\documentclass[12pt,a4paper]{article}
\usepackage{epsfig}
\usepackage[nooneline,flushleft]{caption}
\usepackage{subfigure}
\usepackage{amsmath}
\usepackage{amssymb}
\newcommand{\spp}{\vphantom{\bigg(}}
\newcommand{\ord}{{\cal O}}
\def\gev{{\rm GeV}}
\allowdisplaybreaks
\begin{document}
\title{Lepton flavor violation decays $\tau^-\rightarrow \mu^- P_1 P_2$
in \\the topcolor-assisted technicolor model and \\the littlest
Higgs model with $T$ parity
\\
\hspace*{-0.8cm}  }

\author{Wei Liu, Chong-Xing Yue, Jiao Zhang\\
{\small  Department of Physics, Liaoning Normal University, Dalian
116029, China}\thanks{cxyue@lnnu.edu.cn}\\}
\date{}

\maketitle
\begin{abstract}

The new particles predicted by the topcolor-assisted technicolor
($TC2$) model and the littlest Higgs model with T-parity (called
$LHT$ model) can induce the lepton flavor violation ($LFV$)
couplings at tree level or one loop level, which might generate
large contributions to some $LFV$ processes. Taking into account the
constraints of the experimental data on the relevant free
parameters, we calculate the branching ratios of the $LFV$ decay
processes $\tau^-\to\mu^- P_1 P_2 $ with $P_1 P_2$ = $\pi^+\pi^-$,
$K^+K^-$ and $K^0\bar{K^0}$ in the context of these two kinds of new
physics models. We find that the $TC2$ model and the $LHT$ model can
indeed produce significant contributions to some of these $LFV$
decay processes.
\end{abstract}

\newpage

\noindent{\bf \large 1. Introduction}

In the standard model ($SM$), because of the unitary of the leptonic
analog of Cabibbo-Kobayashi-Maskawa ($CKM$) mixing matrix and the
masslessness of three neutrinos, the lepton flavor violation ($LFV$)
processes are forbidden at tree level. Experimentally, the neutrinos
acquire small mass for the observation of neutrino oscillations and
the $LFV$ processes are possible \cite{1}. Thus, the $LFV$ processes
may provide good tests of new physics ($NP$) beyond the $SM$. This
fact has lead to great amount of the theoretical efforts studying on
the underlying $NP$ in the leptonic flavor sector.

In the $SM$, the $\tau$ lepton is the most heavy particle in the
leptonic sector, which is much more sensitive than the leptons $e$
or $\mu$ to $NP$ related to the flavor and mass generation problems
\cite{2}. The semileptonic $\tau$ decays related $LFV$ are very
interesting and needed to be studied, which could provide a better
laboratory to search $NP$.

Experimentally, with a total data set now exceeding $1.1 ~\rm
ab^{-1}$ of integrated luminosity and a $e^+e^-\to\tau^+\tau^-$
cross-section at $10.58~\rm GeV$ of $0.919 ~\rm nb$ \cite{exp2}, B
factories have recorded more than $10^9$ tau pairs and contributed
significant progress to tau lepton physics. The current experimental
limits for the $LFV$ decay processes $\tau^-\to\mu^- P_1 P_2$ at
$90\%\rm ~C.L.$ have been fixed at \cite{14-belle,155}:
\begin{eqnarray}
Br(\tau^-\to\mu^- \pi^+\pi^-)< 2.9\times10^{-7},\label{1}\\
Br(\tau^-\to\mu^- K^+K^-)< 2.5\times10^{-7}, \label{2}\\
Br(\tau^-\to\mu^- K^0\bar{K^0})<  3.4\times10^{-6}.\label{3}
\end{eqnarray}

There are a lot of theoretical researches on the $LFV$ $\tau$ decays
in many possible extension of the $SM$. For example, the $LFV$
$\tau$ decays have been studied in supersymmetry ($SUSY$) model
\cite{3,301,4,5}, the littlest Higgs model with $T$ parity (called
$LHT$ model) \cite{08,8,88}, and others \cite{6,9,11,12}. In
particular, the $LFV$ $\tau$ decays $\tau\to lP_1P_2$ ($l=\mu, e$)
have been studied in Refs. \cite{3,5,10,tlnp,tlnp2}. However, so
far, we have not found discussions on the $LFV$ $\tau$ decay
processes $\tau^-\to\mu^- P_1 P_2 $ with $P_1 P_2 = \pi^+\pi^-$,
$K^+K^-$ and $K^0\bar{K^0}$ in the framework of the
topcolor-assisted technicolor ($TC2$) model \cite{9-tc2} as well as
the $LHT$ model \cite{7}. These two models are popular and
interesting $NP$ models at present and the experimental upper limits
of the $LFV$ decay processes $\tau^-\to \mu^- P_1 P_2$ have been
improved to $\ord(10^{-7})$ at $90\% ~\rm C.L.$ \cite{14-belle,155}.
So in this paper, we would like to consider the contributions of the
$TC2$ model and the $LHT$ model to the $LFV$ decay processes
$\tau^-\to\mu^- P_1 P_2$.

Among various kinds of dynamical electroweak symmetry breaking
($EWSB$) theories, the topcolor scenario is attractive because it
can explain the large top quark mass and provide a possible $EWSB$
mechanism \cite{15-ewsb}. The $TC2$ model \cite{9-tc2} is one of the
phenomenologically viable models, which has all essential features
of the topcolor scenario. This model predicts the existence of the
nonuniversal gauge boson $Z'$ and the top-Higgs $h^0_{t}$. These new
particles treat the third generation fermions differently from those
in the first and second generations and thus can lead to the tree
level flavor-changing ($FC$) couplings. Thus these new particles
might give significant contributions to the $LFV$ semileptonic
decays $\tau^-\to\mu^-  P_1 P_2$. Our numerical results show that
the contributions of the scalar $h^0_{t} $ are much small, while the
nonuniversal gauge boson $Z'$ can enhance the branching ratio
$Br(\tau^-\to\mu^- P_1 P_2) $ by several orders of magnitude.

The $LHT$ model \cite{7} is one of the attractive little Higgs
models, it predicts the existence of the T-odd $SU(2)$ doublet
fermions and new gauge bosons. These new fermions and gauge bosons
can provide rich phenomenology at present or in future high energy
collider experiments \cite{017,17,18,19,20,201,2x1,21}. Our numerical
results show that the contributions of the $LHT$ model can
significantly enhance the branching ratio $Br(\tau^-\to\mu^- P_1
P_2) $, which might approach its experimental upper limit with
reasonable values of the free parameters.

The structure of this paper is as follows. After briefly summarize
the relevant couplings of new particles to ordinary particles
arising from the $TC2$ model and the $LHT$ model, we calculate the
branching ratios of the $LFV$ decay processes $\tau^-\to\mu^- P_1
P_2 $ with $P_1 P_2$ = $\pi^+\pi^-$, $K^+K^-$ and $K^0\bar{K^0}$
generated by these two kinds of $NP$ models in sections 2 and 3,
respectively. In our numerical estimation, we have taken into
account the constraints of the current experimental data on the
model-dependent free parameters and compared our numerical results
with the current experimental up limits for $\tau^-\to\mu^- P_1 P_2$
in these two sections. Our conclusions and discussions are given in
section 4. In appendix A we give the explicit forms of the relevant
form factors for the pseudoscalar mesons $P_1$ and $P_2$. The
explicit forms of the relevant functions for the $TC2$ and the $LHT$
models are collected in appendixes B and C, respectively.

\vspace{0.4cm}

\noindent{\bf \large 2. The $\rm{TC2}$ model and the $LFV$ $\tau$
decay processes $\tau^-\to \mu^- P_1 P_2$}

In the $TC2$ model \cite{15-ewsb}, topcolor interaction is not
flavor-universal and mainly couples to the third generation
fermions. It generally generates small contributions to $EWSB$ and
gives rise to the main part of the top quark mass. Thus, the
nonuniversal gauge boson $Z'$ has large Yukawa couplings to the
third generation fermions. Such features lead to large tree level
$FC$ couplings of the nonuniversal gauge boson $Z'$ to ordinary
fermions when one writes the interaction in the fermion mass
eigen-basis.

The explicit form for the $LFV$ couplings of the nonuniversal gauge
boson $Z'$ to ordinary leptons, which are related our calculation,
can be written as \cite{17fc,18fc}:
\begin{eqnarray}
L^{FC}_{Z'}=\frac{1}{2}g_{1}K'Z'_{\mu}[\bar{\tau}_{L}\gamma^{\mu}\mu_{L}+
2\bar{\tau}_{R}\gamma^{\mu}\mu_{R}],
\end{eqnarray}
where $g_{1}$ is the ordinary hypercharge gauge coupling constant.
$K'$ is the mixing factor between the leptons $\tau$ and $\mu$. The
relevant flavor-diagonal ($FD$) couplings of $Z'$ to ordinary
fermions can be written as \cite{9-tc2,15-ewsb,17fc}:
\begin{eqnarray}\label{fey-fd}
L^{FD}_{Z'}&=&-\sqrt{4\pi K_{1}}\left\{
Z'_{\mu}\left[\frac{1}{2}\bar{\tau}_{L}\gamma^{\mu}\tau_{L}
-\bar{\tau}_{R}\gamma^{\mu}\tau_{R}\right]-  tan^{2}
\theta'Z'_{\mu}\left[ \frac{1}{6}\bar{u}_{L}\gamma^{\mu}u_{L}+
\frac{2}{3}\bar{u}_{R}\gamma^{\mu}u_{R}\right.\right.\nonumber\\
&&+\left.\left.\frac{1}{6}\bar{d}_{L}\gamma^{\mu}d_{L}-\frac{1}{3}\bar{d}_{R}\gamma^{\mu}d_{R}
+\frac{1}{6}\bar{s}_{L}\gamma^{\mu}s_{L}-\frac{1}{3}\bar{s}_{R}\gamma^{\mu}s_{R}\right]\right\},
\end{eqnarray}
where $K_{1}$ is the coupling constant and $\theta'$ is the mixing
angle with $\tan \theta'={g_{1}}/{\sqrt{4\pi K_{1}}}$.

For the $TC2$ model, the extended gauge groups are broken at the
$TeV$ scale, which proposes that $K'$ is an $O(1)$ free parameter.
Its value can be generally constrained by the present experimental
upper limits on the $LFV$ processes $l_{i}\rightarrow l_{j}\gamma$
and $l_{i}\rightarrow l_{j}l_{k}l_{l}$. For example, for the $LFV$
process $\mu \rightarrow 3e$, the decay width arisen from $Z'$
exchange can be written as \cite{150}:
\begin{eqnarray}
\Gamma (\mu \rightarrow 3e) = \frac{25 \alpha^{5}}{384\pi
K_{1}^{3}cos^{10}\theta_{W}}\frac{m_{\mu}^{5}}{M_{Z'}^{4}}K'^{2},\label{3e}
\end{eqnarray}
where $\theta_{W}$ is the Weinberg angle. The current experimental
upper limit is $Br^{exp}(\mu \rightarrow 3e)\leq 1\times 10^{-12}$
\cite{116}, which can give constraints to the free parameters of the
$TC2$ model. In our following numerical calculation, we will take
into account these limits.

\begin{figure}[htb]
\begin{center}
\epsfig{file=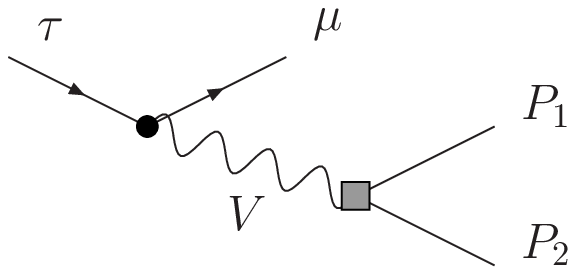,scale=0.95} \caption{The Feynman diagrams
contributing to the $LFV$ decay processes \hspace*{1.8cm}$\tau^-\to
\mu^- P_1 P_2$. $V$ represents a photon, a gauge boson or a Higgs
boson.}\label{fe1}
\end{center}
\end{figure}


From the above discussions, we can see that the nonuniversal gauge
boson $Z'$ can contribute to the $LFV$ decay processes $\tau^-\to
\mu^- P_1 P_2$ at tree level and one loop level as shown in
Fig.~\ref{fe1}. This diagram can be mediated by a photon, a gauge
boson or a Higgs boson. Here the effective $LFV$ vertex is
represented by a black dot and the hadronic vertex by a gray box.
There are also other types of diagrams induced by the gauge bosons
$W^{\pm}$ which have been discussed in Ref. \cite{10}. However,
those diagrams do not exist in our calculation, because they are
just adapt to models including right-handed neutrinos. The
pseudoscalar mesons $P_1$ and $P_2$ in the final state stem from the
hadronisation of quark bilinear currents, namely parameterizing by
the vector form factors $F^{P_1P_2}(s)$ \cite{4,51form}. These form
factors can be defined through the vacuum-to-$P_1P_2$ matrix
elements of the local quark currents. The relevant formula can be
written as \cite{4,51form}:
\begin{eqnarray}\label{form}
\langle P_1P_2|\bar{q}\gamma_{\mu}q|0\rangle
=(p_1-p_2)_{\mu}F_q^{P_1P_2}(s) ,
\end{eqnarray}
and $\sum_q^{u,d,s} Q_qF_q^{P_1P_2}(s)=F^{P_1P_2}(s)$, where $Q_q$
is the electric charge of the $q$ quark in units of the positron
charge $e$ and $s=(p_1+p_2)^2$, in which $p_1$ and $p_2$ are the
momentum of mesons $P_1$ and $P_2$, respectively. The explicit forms
of $F^{\pi^+\pi^-}(s)$, $F^{K^+K^-}(s)$ and $F^{K^0\bar {K^0}}(s)$
have been displayed in the appendix A.

In this section, we give the explicit calculation of the $Z'$
contributions to the $LFV$ decay processes $\tau^-\to \mu^- P_1 P_2$
at both tree level and one loop level.

\vspace{0.4cm}

\noindent{ \bf A. The tree level contributions of the nonuniversal
gauge boson $Z'$}

From Eq.~(4), we can see that the nonuniversal gauge boson $Z'$ can
contribute to the $LFV$ decay processes $\tau^-\to \mu^- P_1 P_2$ at
tree level. The relevant Feynman diagram is similar to
Fig.~\ref{fe1}. The amplitude mediated by  $Z'$ exchange in terms of
the final state quarks can be written as:
\begin{equation}\label{a}
A_{Z'}=\frac{1}{M_{Z'}^2}C_{Z'}~\bar\mu
\gamma_{\mu}(v_l+a_l\gamma_5)\tau ~\bar{q}\gamma_{\nu}
(v_q+a_q\gamma_5)q,
\end{equation}
in which $v_{l(q)}$ and $a_{l(q)}$ are the constants for the vector
and axial-vector couplings of the gauge boson $Z'$ to ordinary
leptons (quarks). The coefficient $C_{Z'}$ can be written as:
\begin{equation}\label{b}
C_{Z'}=\frac{1}{2} g_1K'\sqrt{4\pi K_1}tan^2\theta'.
\end{equation}

Utilizing the hadronisation formula given by Eq.~(\ref{form}), the
quark bilinear currents can be written in term of the form factors
$F_q^{P_1P_2}(s)$ which correspond to the two mesons $P_1$ and $P_2
$ in the final state. Then, in terms of the final state hadrons, we
can obtain the amplitude of the $LFV$ decay processes $\tau^-\to
\mu^- P_1 P_2$ generated by the nonuniversal gauge boson $Z'$
\begin{equation}\label{c}
A_{Z'}=\frac{v_q}{M_{Z'}^2}C_{Z'}F_q^{P_1P_2}(s)~\bar\mu
(p_1\!\!\!\!/-p_2\!\!\!\!/)(v_l+a_l\gamma_5)\tau.
\end{equation}
The explicit form of the branching ratio $ Br(\tau^-\to \mu^- P_1
P_2) $ can be expressed as \cite{4}:
\begin{equation}\label{c}
Br(\tau^-\to \mu^- P_1
P_2)=\frac{\tau_\tau}{64\pi^3m_{\tau}^2}\int_{s_{min}}^{s_{max}}ds\int_{t_{min}}^{t_{max}}dt
~|A_{Z'}|^2,\
\end{equation}
where $\tau_{\tau}$ is the lifetime of lepton $\tau$,
$t=(p_{\tau}-p_1)^2$, and
\begin{eqnarray}
t_{\rm min}^{\rm max} &= &  \frac{1}{4 s} \left[ \left( m_\tau^2 -
m_\mu^2 \right)^2 - \left( \lambda^{1/2}\left( s,m_{P}^2,
m_{P}^2 \right) \mp \lambda^{1/2} \left( m_\tau^2, s, m_\mu^2
\right)\right)^2 \right] , \nonumber \\
s_{\rm min}  &=&  4 m_P^2 \,\,,\,\, s_{\rm max} =  \left( m_\tau -
m_\mu \right)^2 \, \,, \, \, \lambda(x,y,z) = (x+y-z)^2-4xy \,.
\label{lambda}
\end{eqnarray}
In above equations we have assumed $m_{P_1}=m_{P_2}=m_{P}$.

\begin{table}
\begin{center}
\begin{displaymath}
\begin{tabular}{|l|l|}
\hline \spp $G_F = 1.166 \times 10^{-5} \; \gev^{-2}$ &
$m_{\tau}=1.78 \; \gev$ \\
\spp $\alpha = 7.297 \times 10^{-3}$ &
$ m_{\mu}=0.106 \; \gev$  \\
\spp $\tau_{\tau} = 2.91 \times 10^{-13} s$ &
$m_{K}= 0.494 \; \gev$  \\
\spp $m_{\pi}= 0.139 \; \gev$  &
$m_{K^0}= 0.498 \; \rm GeV $ \\
\spp $M_W= 80.43 \; \gev$ & $sin^2\theta_W=0.2315$
\\ \hline
\end{tabular}
\end{displaymath}
\caption{Numerical inputs used in our analysis. Unless explicitly
specified, they are \hspace*{1.7cm}taken from the Particle Data
Group \cite{155}.\label{tab:inputs}}
\end{center}
\end{table}

Before giving numerical results, we need to specify the relevant
$SM$  parameters. Most of these input parameters are shown in Table
\ref{tab:inputs}. The vacuum tilting, the constraints from Z-pole
physics, and U(1) triviality require $K_{1}\leq 1$ \cite{1t8}. The mass of
nonuniversal gauge boson $M_{Z'}$ can be generally seen as free
parameter. The lower bounds on $M_{Z'}$ can be obtained from dijet
and dilepton production in the Tevatron experiments \cite{125} or
$B\bar{B}$ mixing \cite{126}. However, these bounds are
significantly weaker than those from the precision electroweak data.
Ref. \cite{160} has shown that, to fit the precision electroweak
data, the $Z'$ mass $M_{Z'}$ must be larger than $1~\rm TeV$. In the
following numerical estimation, we will assume that the values of
the free parameters $M_{Z'}$ and $K_{1}$ are in the ranges of $1000\
\rm GeV\ \sim 2000\ \rm GeV$ and $0 \sim 1$, respectively.

\begin{figure}[htb]
\begin{center}
\subfigure[$K_1=0.4$]{
\includegraphics[scale=0.85]{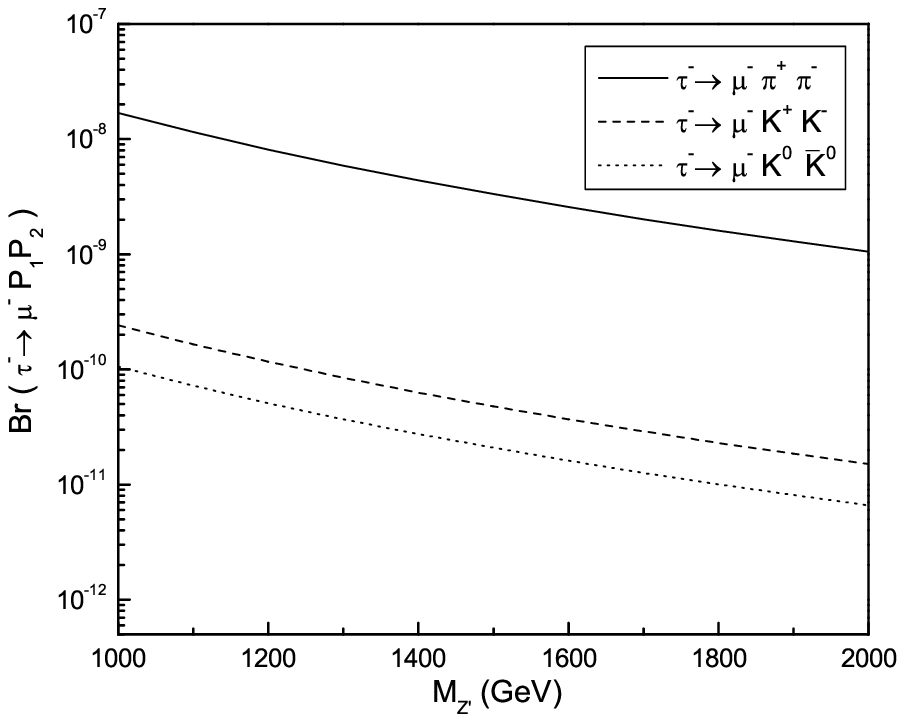}}
\subfigure[$K_1=0.8$]{
\includegraphics[scale=0.85]{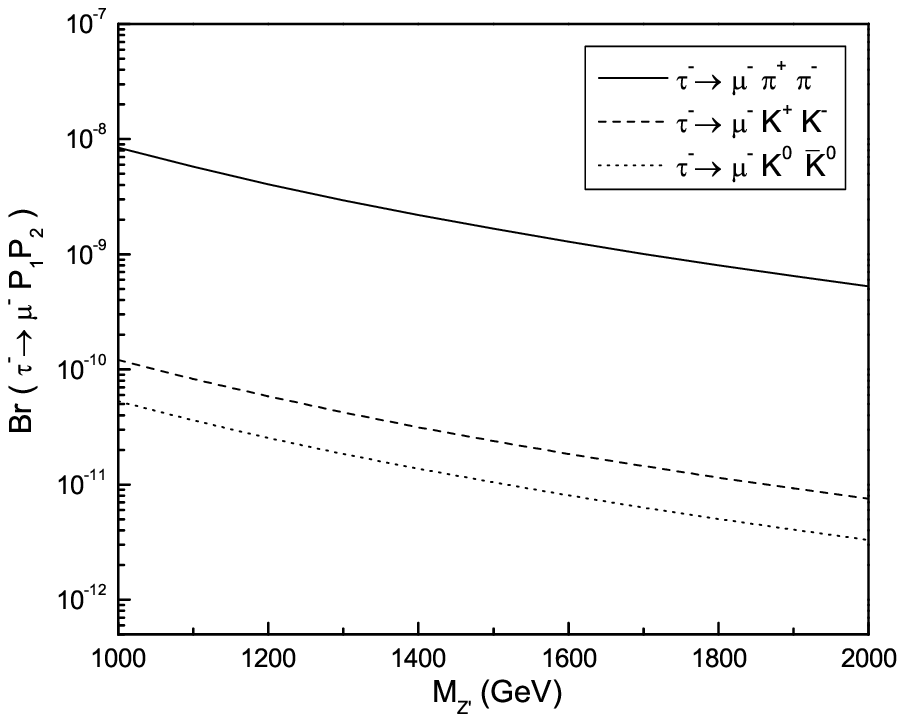}}
\caption{
The branching ratios $Br(\tau^-\to \mu^- P_1 P_2)$ contributed by the nonuniversal \hspace*{1.8cm}
gauge boson $Z'$ at tree level as functions of mass parameter $M_{Z'}$ for the \hspace*{1.8cm}
parameter $K_{1}=0.4$ (a) and $K_{1}=0.8$ (b).} \label{fi1}
\end{center}
\end{figure}

The branching ratios $Br(\tau^-\to \mu^- P_1 P_2)$ with
$P_1P_2=\pi^+\pi^-$, $K^+K^- $ and $K^0\bar {K^0}$ contributed by
the nonuniversal gauge boson $Z'$ at tree level are plotted as
functions of the mass parameter $M_{Z'}$ in Fig.~\ref{fi1}, in which
we have taken $K_{1}=0.4$ (Fig.~\ref{fi1}a) and 0.8(Fig.~\ref{fi1}b), and considered
the constraints on the free parameter $K'$ giving by the current
experimental upper limit of $Br^{exp}(\mu \rightarrow 3e)$, as shown
in Eq.~(\ref{3e}). From these diagrams we can see that the values of
the branching ratios $Br(\tau^-\to \mu^- \pi^+\pi^-)$, $Br(\tau^-\to
\mu^- K^+K^-)$ and $Br(\tau^-\to \mu^- K^0\bar {K^0})$ decrease as
the mass parameter $M_{Z'}$ increasing. It is obviously that the
branching ratios of the different decay channels satisfy the
relation  $Br(\tau^- \to \mu^- \pi^+ \pi^-)$ $>$ $BR(\tau^- \to
\mu^- K^+ K^-)$ $\gtrsim$ $BR(\tau^- \to \mu^- K^0 \bar{K}^0)$. This
is mainly because the mass of the $K$ meson is larger than that of
the $\pi$ meson and the $FD$ couplings of the nonuniversal gauge
boson $Z'$ to up-type quarks are different from those for the
down-type quarks as shown in Eq.~(\ref{fey-fd}). The max values of
the branching ratios for the $LFV$ decay processes $\tau^-\to \mu^-
K^+K^-$ and $\tau^-\to \mu^- K^0\bar {K^0}$ can reach $2.41\times
10^{-10}$ and $1.05\times10^{-10}$, respectively. However, these
values are much smaller than the corresponding experimental upper
limits given in Eqs.~(\ref{2}, \ref{3}). While the max value for the
$LFV$ decay process $\tau^-\to \mu^- \pi^+\pi^-$ can reach
$1.68\times 10^{-8}$, which might approach its upper limit in the
future high energy collider experiments.

\vspace{0.4cm}

\noindent{ \bf B. The loop level contributions of the nonuniversal
gauge boson $Z'$}

\begin{figure}[htb]
\begin{center}
\epsfig{file=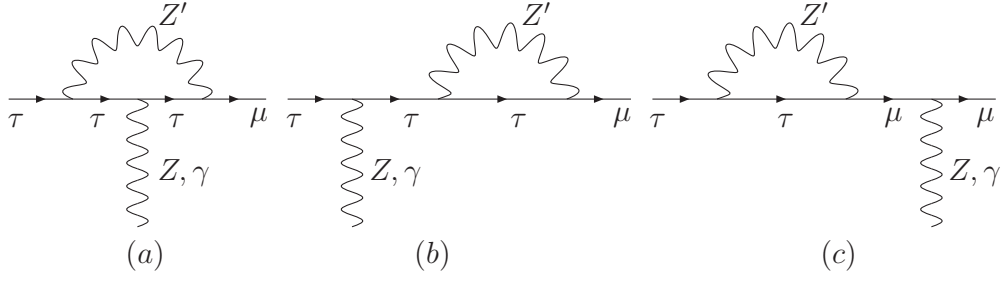,scale=0.8}
\caption{The Feynman diagrams for the effective $LFV$ vertexes $Z \tau  \bar{\mu}$ and
$\gamma \tau \bar{\mu} $ \hspace*{1.8cm}contributed  by the nonuniversal
gauge boson $Z'$.}
\label{fe2}
\end{center}
\end{figure}

The nonuniversal gauge boson $Z'$ predicted by the $TC2$ model can
also generate contributions to the $LFV$ decay processes $\tau^-\to
\mu^- P_1 P_2$ at one loop level. The relevant Feynman diagrams for
the effective $LFV$ vertexes $Z \tau  \bar{\mu}$ and $\gamma \tau
\bar{\mu} $ have been displayed in Fig.~\ref{fe2}.

The effective Hamilton for the $LFV$ decay process $\tau^-\to \mu^-
P_1 P_2$ including the contributions of $Z'$ at one loop level has
the form:
\begin{eqnarray}
H&=&H_1+H_2, \\
H_1&=&\frac{G_F}{\sqrt{2}}\frac{\alpha}{2\pi
sin^2\theta_W}C_1~\bar\mu\gamma_{\mu}(v_l+a_l\gamma_5)\tau
~\bar{q}\gamma_{\nu}(v_q+a_q\gamma_5)q , \\
H_2&=&\frac{G_F}{\sqrt{2}}\frac{m_{\tau}e^2Q_q}{4\pi^2k^2}C_2
~\bar{\mu}i\sigma_{\mu\nu}(v_l+a_l\gamma_5)\tau~\bar{q}\gamma_{\nu}q,
\end{eqnarray}
where $H_1$ and $H_2$ represent the $Z'$ contributions mediated by
$Z$ gauge boson exchange and the photon exchange, respectively. $k$
represents the photon momentum. The explicit forms of the
coefficients $C_1$ and $C_2$ are:
\begin{eqnarray}
C_1 &=& \frac{2g_1K'\sqrt{4\pi K_1}}{g_2^2}\left[4F_1(x_{\tau})
-2F_2(x_{\tau})+\left(1+\frac{m_{\tau}}{m_{\mu}}F_3(x_{\tau})\right)\right] ,\\
C_2 &=& \frac{32g_1K'M_W^2\sqrt{4\pi K_1}}{g_2^2
m_{\tau}}\left[F_4(x_{\tau})+\left(1+\frac{m_{\tau}}{m_{\mu}}F_3(x_{\tau})\right)\right],
\end{eqnarray}
where $g_2$ is the $SM$ $SU(2)_{L}$ gauge coupling constant. The
Inami-Lim functions \cite{smf} $F_i(x)$ ($i=1,2,3,4$) are collected
in Appendix B with $x_{\tau}={m_{\tau}^2}/{M_{Z'}^2}$.

Applying similar hadronisation process to the bilinear quark
currents as that for the tree level, the amplitude contributed by
the nonuniversal gauge boson $Z'$ at one loop can be written as:
\begin{eqnarray}
A_{1}&=&\frac{G_F}{\sqrt{2}}\frac{v_q\alpha}{2\pi
sin^2\theta_W}C_{1}F_q^{P_1P_2}(s)~\bar\mu
(p_1\!\!\!\!/-p_2\!\!\!\!/)(v_l+a_l\gamma_5)\tau,\\
A_{2}&=&\frac{G_F}{\sqrt{2}}\frac{e^2 m_{\tau}}{2\pi^2 k^2
}C_{2}F^{P_1P_2}(s)~\bar\mu
p_1^{\mu}\sigma_{\mu\nu}p_2^{\nu}(v_l+a_l\gamma_5)\tau.
\end{eqnarray}

In the context of the $TC2$ model, the expression of the
corresponding branching ratio induced by the nonuniversal gauge
boson $Z'$ at one loop level can be written as:
\begin{equation}\label{f}
Br(\tau^-\to \mu^- P_1
P_2)=\frac{\tau_\tau}{64\pi^3m_{\tau}^2}\int_{s_{min}}^{s_{max}}ds\int_{t_{min}}^{t_{max}}dt
~\left(|A_{1}|^2+|A_{2}|^2\right).
\end{equation}

\begin{figure}[htb]
\begin{center}
\subfigure[$K_1=0.4$]{
\includegraphics[scale=0.85]{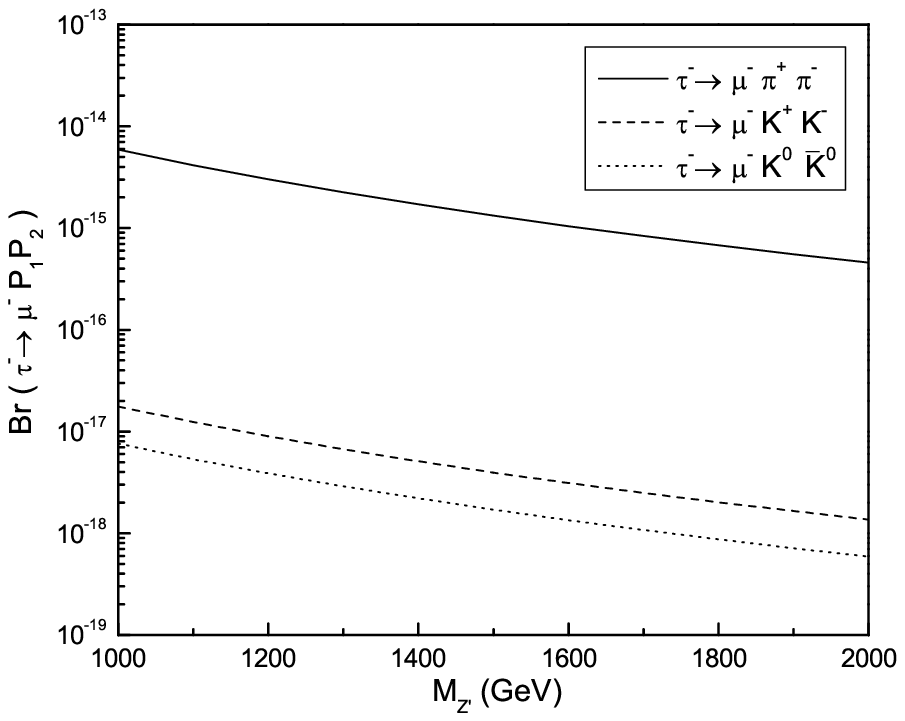}}
\subfigure[$K_1=0.8$]{
\includegraphics[scale=0.85]{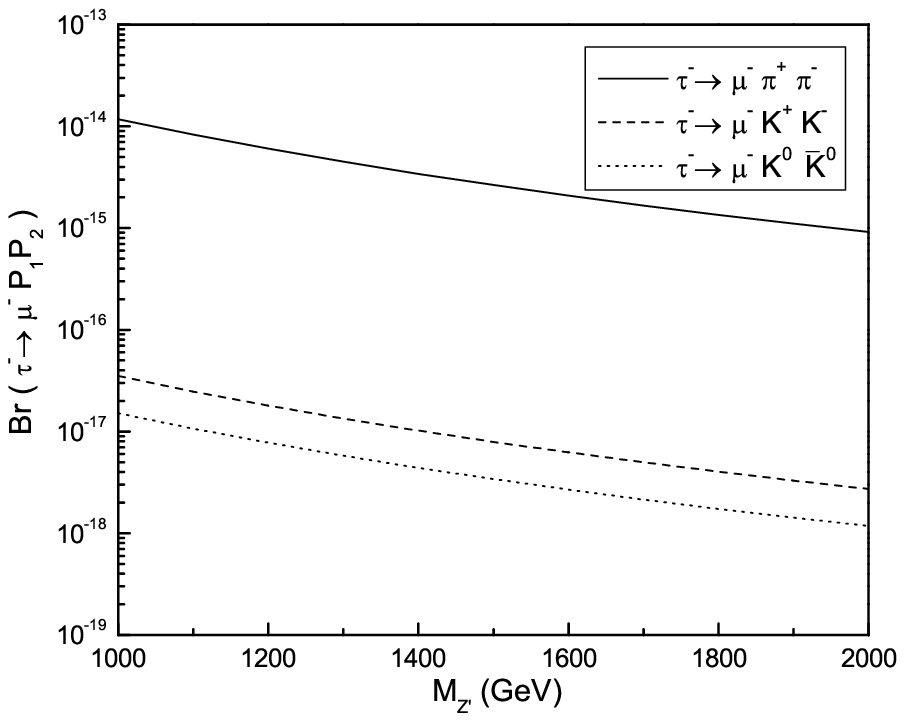}}
\caption{The branching ratios
$Br(\tau^-\to \mu^- P_1 P_2)$ as functions of mass parameter
$M_{Z'}$ \hspace*{1.8cm} at one loop level for the parameter $K_{1}=0.4$ (a) and $K_{1}=0.8$ (b).} \label{fi3}
\end{center}
\end{figure}

Using the values of the relevant $SM$ input parameters given at
Table \ref{tab:inputs}, we present the branching ratios
$Br(\tau^-\to \mu^- \pi^+\pi^-)$, $Br(\tau^-\to \mu^- K^+K^-)$ and
$Br(\tau^-\to \mu^- K^0\bar {K^0})$ contributed by the nonuniversal
gauge boson $Z'$ at one loop level as functions of the mass
parameter $M_{Z'}$ in Fig.~\ref{fi3}, in which
we have taken $K_{1}=0.4$ (Fig.~\ref{fi3}a) and 0.8(Fig.~\ref{fi3}b).
From these diagrams, one can see
that the values of the branching ratios $Br(\tau^-\to \mu^-
\pi^+\pi^-)$, $Br(\tau^-\to \mu^- K^+K^-)$ and $Br(\tau^-\to \mu^-
K^0\bar {K^0})$ decrease as the mass parameter $M_{Z'}$ increasing.
The value of $Br(\tau^-\to \mu^- K^+K^-)$ is close to that of
$Br(\tau^-\to \mu^- K^0\bar {K^0})$ in most of the parameter space
of the $TC2$ model. Comparing this figure to the $Z'$ tree level
contributions displayed in Fig.~\ref{fi1}, one can see that the
contributions of $Z'$ to the $LFV$ decay processes $\tau^-\to \mu^-
\pi^+\pi^-$, $\tau^-\to \mu^- K^+K^-$ and $\tau^-\to \mu^- K^0\bar
{K^0}$ at one loop level are smaller than those of the tree level
diagram by several orders of magnitude in most of the parameter space.

The $TC2$ model also predicts the existence of the top-Higgs
$h^0_{t}$, which treats the third generation fermions differently
from those in the first and second generations and thus can lead to
the tree level $FC$ couplings to ordinary fermions. So this kind of
new particle can also generate contributions to the $LFV$
semileptonic decays $\tau^-\to\mu^-  P_1 P_2$ at tree level and one
loop level. However, the $LFV$ coupling $h_t^0\tau\mu $ is
suppressed by a factor ${m_{\tau}}/{\nu}$ with the electroweak scale
$\nu=246~\rm GeV$. Thus, the contributions of the top-Higgs
$h^0_{t}$ to the $LFV$ semileptonic decays $\tau^-\to\mu^-  P_1 P_2$
are much smaller than those of the nonuniversal gauge boson $Z'$.
Our numerical results show that it indeed is this case. The value of
the branching ratio $Br(\tau^-\to \mu^- P_{1}P_{2})$ contributed by
the scalar $h^0_{t} $ is smaller than that of $Z'$ at least by two
orders of magnitude.

\vspace{0.4cm}

\noindent{\bf \large 3. The $\rm{LHT}$ model and the $LFV$ $\tau$
decay process $\tau^-\to \mu^- P_1 P_2$}

In this section, we first review the essential features of the $LHT$
model studied in Ref. \cite{7}, which are related our calculation.
Then we will consider the contributions of the $LHT$ model to  the
$LFV$ $\tau$ decay process $\tau^-\to \mu^- P_1 P_2$.

Similar with the $LH$ model, the $LHT$ model is based on an
$SU(5)/SO(5)$ global symmetry breaking pattern. A subgroup
$[SU(2)\times U(1)]_{1}\times [SU(2)\times U(1)]_{2}$ of the $SU(5)$
global symmetry is gauged, and at the scale $f$ it is broken into
the $SM$ electroweak symmetry $SU(2)_{L}\times U(1)_{Y}$. T-parity
is an automorphism which exchanges the $[SU(2)\times U(1)]_{1}$ and
$[SU(2)\times U(1)]_{2}$ gauge symmetries. The T-even combinations
of the gauge fields are the $SM$ electroweak gauge bosons
$W_{\mu}^{a}$ and $A_{\mu}$. The T-odd combinations are T-parity
partners of the $SM$ electroweak gauge bosons.

After taking into account $EWSB$, at the order of $\nu^{2}/f^{2}$,
the masses of the T-odd set of the $SU(2)\times U(1)$ gauge bosons
are given as:
\begin{equation}
M_{A_{H}}=\frac{g_1f}{\sqrt{5}}
\left[1-\frac{5\nu^{2}}{8f^{2}}\right], \hspace{0.5cm}M_{Z_{H}}
\approx M_{W_{H}}=g_2f\left[1-\frac{\nu^{2}}{8f^{2}}\right],
\end{equation}
where $f$ is the scale parameter of the gauge symmetry breaking of
the $LHT$ model. Because of the smallness of $g_1$, the T-odd gauge
boson $A_{H}$ is the lightest T-odd particle, which can be seen as
an attractive dark matter candidate \cite{017,16dark}.

To avoid severe constraints and simultaneously implement T-parity,
it is need to double the $SM$ fermion doublet spectrum \cite{7,17}.
The T-even combination is associated with the $SU(2)_{L}$ doublet,
while the T-odd combination is its T-parity partner. The masses of
the T-odd fermions can be written in a unified manner as:
\begin{equation}
M_{F_{i}}=\sqrt{2}k_{i}f,
\end{equation}
where $k_{i}$ are the eigenvalues of the mass matrix $k$ and their
values are generally dependent on the fermion species $i$.

The mirror fermions (T-odd quarks and T-odd leptons) have new flavor
violating interactions with the $SM$ fermions mediated by the new
gauge bosons $(A_{H},W_{H}^{\pm}$, or $Z_{H})$, which are
parameterized by four $CKM$-$like$ unitary mixing matrices, two for
mirror quarks and two for mirror leptons \cite{20,201,18mir}:
\begin{equation}
V_{Hu},\hspace*{0.2cm}V_{Hd},\hspace*{0.2cm}V_{Hl},\hspace*{0.2cm}V_{H\nu},
\end{equation}
they satisfy:
\begin{equation}
V_{Hu}^{+}V_{Hd}=V_{CKM},\hspace*{0.2cm}V_{H\nu}^{+}V_{Hl}=V_{PMNS},
\end{equation}
where the $CKM$ matrix $V_{CKM} $ is defined through flavor mixing
in the down-type quark sector, while the $PMNS$ matrix $V_{PMNS} $
is defined through neutrino mixing. Similar with Ref. \cite{201}, we
will set the Majorana phases of $V_{PMNS}$ to zero in our following
calculation. The matrix $V_{Hl}$ can give rise to the $LFV$
processes.

\begin{figure}[htb]
\begin{center}
\epsfig{file=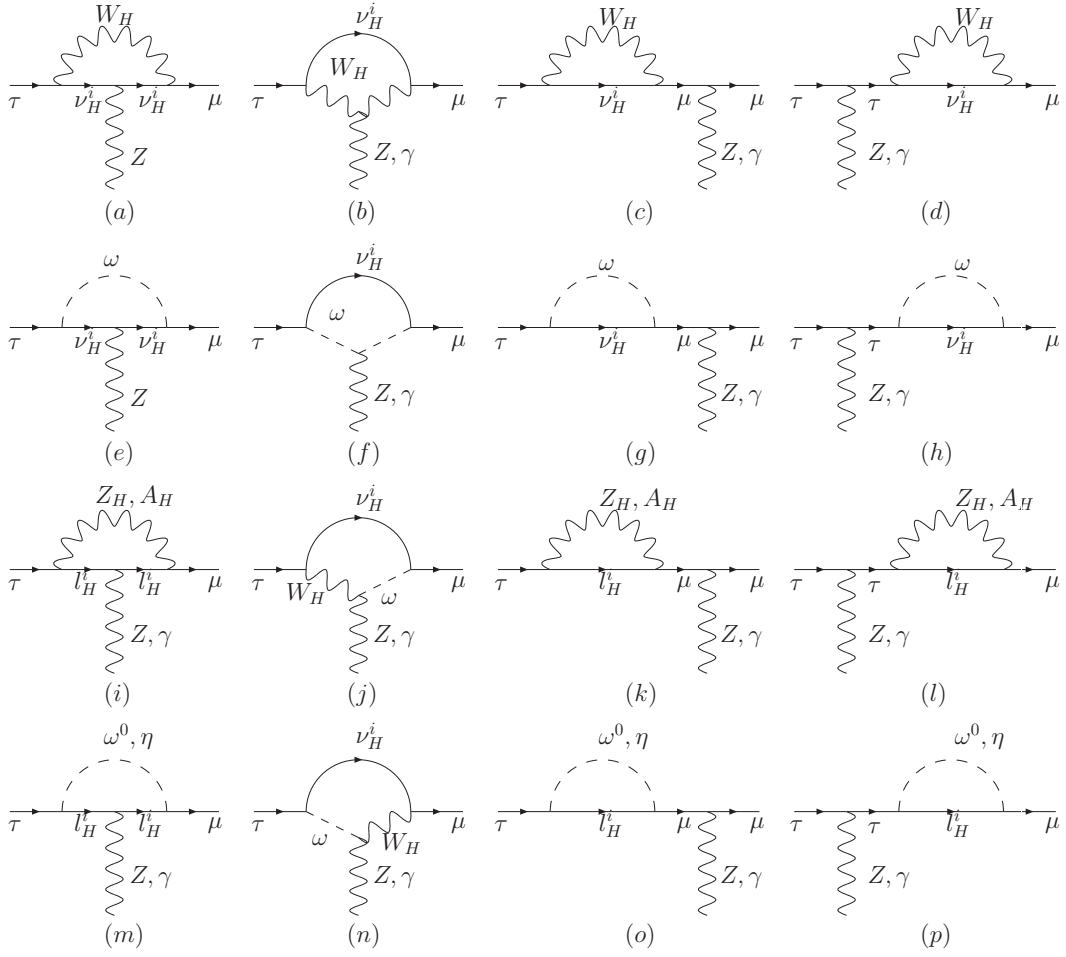,scale=0.65} \caption{The penguin diagrams for
the effective $LFV$ vertexes $Z \tau  \bar{\mu}$ and $\gamma \tau
\bar{\mu} $ in the \hspace*{1.8cm}$ LHT $ model.} \label{fe4}
\end{center}
\end{figure}

\begin{figure}[htb]
\begin{center}
\epsfig{file=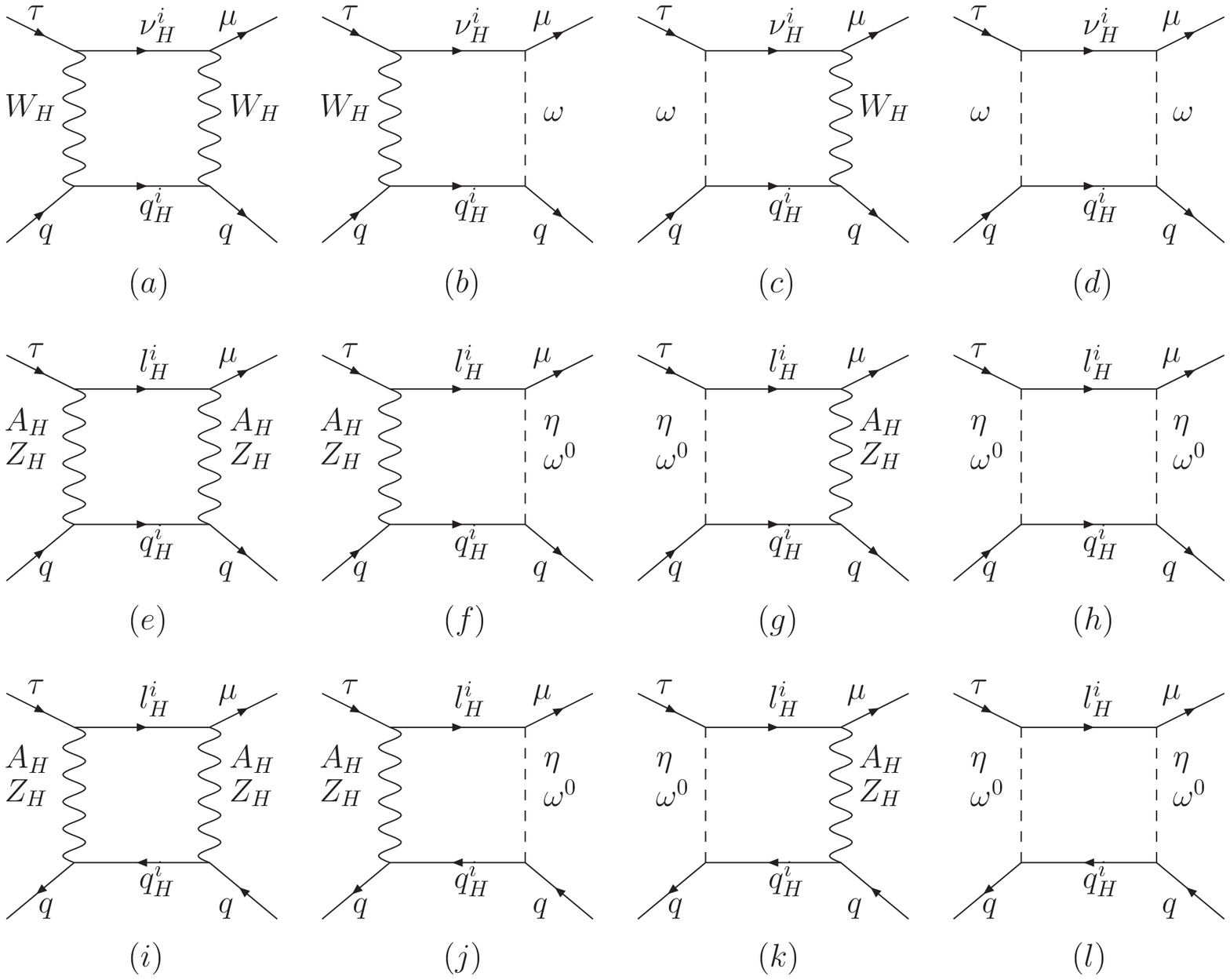,scale=0.75} \caption{The box diagrams for the
$LFV$ decay process $\tau^-\to \mu^- P_1 P_2$ in the $LHT$
\hspace*{1.8cm} model.} \label{fe5}
\end{center}
\end{figure}

From the above discussions, we can see that the $LHT$ model provides
a new mechanism for the $LFV$ processes, which comes from the flavor
mixing in the mirror lepton sector. Thus, the $LHT$ model might give
significant contributions to the $LFV$ processes $\tau^-\to \mu^-
P_1 P_2$. The relevant Feynman diagrams have been shown in
Fig.~\ref{fe4} and Fig.~\ref{fe5}, in which we just display the
effective $LFV$ vertex without hadronic part. In these diagrams,
$l_H^{i}$, $\nu_H^{j}$ and $q_H^i$ represent the T-odd partners of
three family leptons $l_i$, $\nu_j$ and quarks $q_i$, respectively.
The Goldstone bosons $\omega^{\pm}$, $\omega^0$ and $\eta$ are eaten
by heavy gauge bosons $W_H^{\pm}$, $Z_H$ and $A_H$, respectively.
In this paper we use the 't Hooft-Feynman gauge, so the Goldstone
Boson mass is the same as its corresponding gauge boson, that's to say: $M_{\omega}=M_{W_H}$,
$M_{\omega^0}=M_{Z_H}$ and $M_{\eta}=M_{A_H}$. The relevant couplings of these new particles to
ordinary leptons and their T-odd partners can be found in Ref.
\cite{201}. The effective Hamilton for the $LFV$ decay process
$\tau^-\to \mu^- P_1 P_2$ can be written as \cite{201}:
\begin{eqnarray}
H_3&=&\frac{G_F}{\sqrt{2}}\frac{\alpha}{2\pi
sin^2\theta_W}\bar{X}_\text{odd}~\bar\mu\gamma_{\mu}(1-\gamma_5)\tau
~\bar{q}\gamma_{\nu}(v_q+a_q\gamma_5)q  ,\\
H_4&=&\frac{G_F}{\sqrt{2}}\frac{ e^2
m_{\tau}Q_q}{4\pi^2k^2}\bar{D}_\text{odd}
~\bar{\mu}i\sigma_{\mu\nu}k^{\nu}(1+\gamma_5)\tau~\bar{q}\gamma_{\nu}q,
\end{eqnarray}
with
\begin{eqnarray}
\bar{X}_\text{odd}^{u}&=& \left[ \chi_2^{(\tau \mu)}\big(J^{u\bar
u}(y_2,z)-J^{u\bar u}(y_1,z)\big) +\chi_3^{(\tau \mu)}\big(J^{u \bar
u}(y_3,z)-J^{u \bar u}(y_1,z)\big) \right],
\label{Xodd}\\
\bar{X}_\text{odd}^{d}&=& \left[ \chi_2^{(\tau \mu)}\big(J^{d\bar
d}(y_2,z)-J^{d \bar d}(y_1,z)\big) +\chi_3^{(\tau \mu)}\big(J^{d\bar
d}(y_3,z)-J^{d\bar d}(y_1,z)\big) \right] ,\label{Yodd}\\
\bar{D}_\text{odd} &=& -\frac{\nu^2}{8f^2}\sum_{i}
\chi_i^{\tau\mu}\left[D'_0(y_{i})-\frac{7}{6}E_0'(y_{i})-\frac{1}{10}E_0'(y_i')\right],
\end{eqnarray}
here
\begin{eqnarray}
J^{u \bar{u}}\left(y_{i}, z\right)&=&
\frac{1}{64}\frac{v^2}{f^2}\bigg[y_i S_\text{odd}(y_i) +F^{u
\bar{u}}(y_i,z;W_H)\nonumber\\
&& \qquad +4
\Big(G(y_i,z;Z_H)+G_1(y'_i,z';A_H)-G_2(y_i,z;\eta)\Big)\bigg],
\label{Znunu}\\
J^{d\bar d}\left(y_{i}, z\right)&=&
\frac{1}{64}\frac{v^2}{f^2}\bigg[{y_i} S_\text{odd}(y_i) +F^{d\bar d}(y_i,z;W_H)\nonumber\\
&& \qquad -4
\Big(G(y_i,z;Z_H)+G_1(y'_i,z';A_H)+G_2(y_i,z;\eta)\Big)\bigg]
,\label{Zmumu}
\end{eqnarray}
where $y_i={M_{l_H^i}^2}/{M_{W_H}^2}={M_{l_H^i}^2}/{M_{Z_H}^2}$, $z
= {{m^2_{q_{H}}}}/{M_{W_H}^2}$, $y_i(z)'={5}y_i(z)/{tan^2\theta_W}$,
$\eta = {tan^2\theta_W}/{5}$ and
$\chi_i^{\tau\mu}=V_{Hl}^{*i\mu}V_{Hl}^{i\tau}$. The explicit forms
of $S_{odd}(x)$, $F^{u,d}(x)$, $G_i(x)$, $D_0'(x)$ and $E_0'(x)$ are
collected in Appendix C.

In the context of the $LHT$ model, the amplitude of the $LFV$ decay
process $\tau^-\to \mu^- P_1 P_2$ can be written as:
\begin{eqnarray}
A_3&=&\frac{G_F}{\sqrt{2}}\frac{v_q\alpha}{2\pi
sin^2\theta_W}F_q^{P_1P_2}(s)\bar{X}_\text{odd}~\bar\mu(p_1\!\!\!\!/-p_2\!\!\!\!/)(1-\gamma_5)\tau,\\
A_4&=&\frac{G_F}{\sqrt{2}}\frac{ e^2
m_{\tau}}{2\pi^2k^2}F^{P_1P_2}(s)\bar{D}_\text{odd}
~\bar{\mu}ip_1^{\mu}\sigma_{\mu\nu}p_2^{\nu}(1+\gamma_5)\tau.
\end{eqnarray}

The contributions of the $LHT$ model to $LFV$ decay process have been extensively studied
and compared with the current experimental limits in the literatures \cite{8,88,201,2x1}. It has been shown that the
$LHT$ model can enhance the $SM$ prediction values by several orders of magnitude and the experimental
measurement data for some $LFV$ decay processes can give constraints on the free parameters of the $LHT$
model. For example, in order to suppress the branching ratio $Br(\mu\to e\gamma)$ and $Br(\mu\to3e)$
predicted by the $LHT$ model below the present experimental upper bounds, the relevant mixing matrix
$V_{Hl}$ must be rather hierarchical or mass splitting for the first and second T-odd lepton masses is
very small. Ref.\cite{201} has shown that there must be $sin2\theta\leq0.05$ or $\delta\leq5\%$. A complete
analysis can be found in Ref.\cite{201}. Thus, in our following numerical estimation, we will assume
$M_{l^{e}_H}=M_{{\nu}^{e}_H}=M_{l^{\mu}_H}=M_{{\nu}^{\mu}_H}=M_1=800~\rm GeV$, $V_{Hl}=V^+_{PMNS}$, and take
 $M_{l^{\tau}_H}=M_{{\nu}^{\tau}_H}=M_2$ and the scale parameter $f$ as free parameters. Considering the
mirror quarks only contribute to the branching ratios of decay
$\tau^-\to \mu^- P_1 P_2$ in box diagrams, we assume their masses
degeneration and take $M_{q_H}= 1~\rm TeV$.

\begin{figure}
\centering \subfigure[$M_2=500~\rm GeV$]{
\includegraphics[scale=0.85]{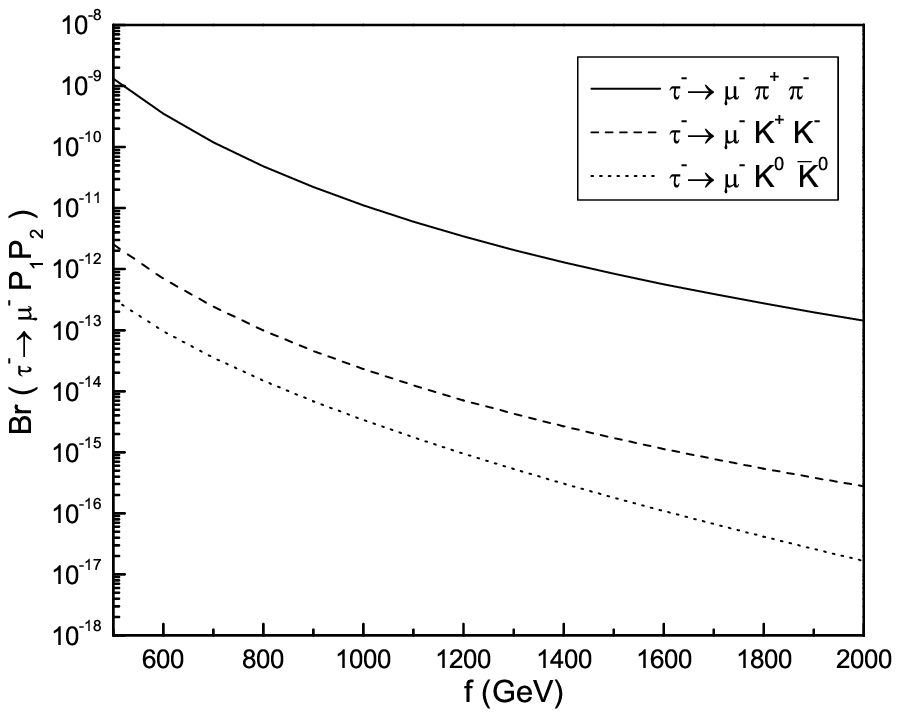}}
\subfigure[$M_2=1500~\rm GeV$]{
\includegraphics[scale=0.85]{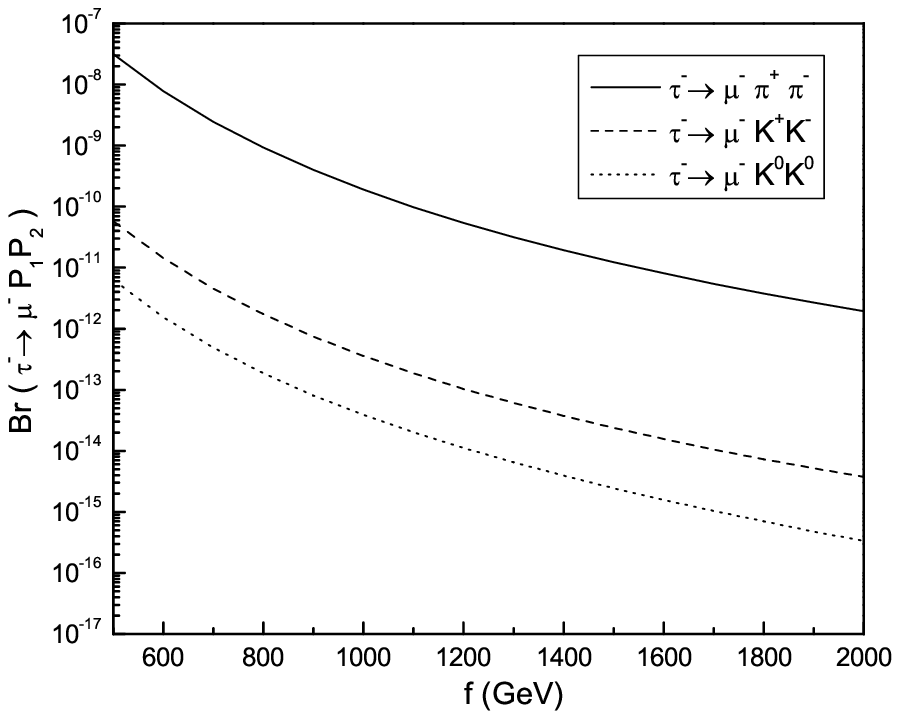}}
\caption{ In the $LHT$ model, the branching ratio $Br(\tau^-\to
\mu^- P_1 P_2)$ as function of $f$ \hspace*{1.8cm}for the parameter
$M_2=500~\rm GeV$ (a), $M_2=1500~\rm GeV$ (b).}
\label{fi46}
\end{figure}

The branching ratios $Br(\tau^-\to \mu^- P_1 P_2)$ with
$P_1P_2=\pi^+\pi^-$, $K^+K^- $ and $K^0\bar {K^0}$ contributed by
the $LHT$ model are plotted as functions of the scale parameter $f$
for $M_2=500~\rm GeV$ (Fig.~\ref{fi46} (a)) and
$M_2=1500~\rm GeV$ (Fig.~\ref{fi46} (b)). From these figures,
one can see that the values of the branching ratios $Br(\tau^-\to
\mu^- \pi^+\pi^-)$, $Br(\tau^-\to \mu^- K^+K^-)$ and $Br(\tau^-\to
\mu^- K^0\bar {K^0})$ decrease as the scale parameter $f$ increasing
while as the mass of T-odd lepton $M_{l^{\tau}_H}$ decreasing. For
$M_{l^{\tau}_H}=1500~\rm GeV$ and $f=500~\rm GeV$, the value of the
branching ratio $Br(\tau^-\to \mu^- \pi^+\pi^-)$ can reach
$3.14\times 10^{-8}$, which is larger than that induced by the $TC2$
model. For the $LFV$ decay processes $\tau^-\to \mu^- K^+K^-$ and
$\tau^-\to \mu^- K^0\bar {K^0}$, the values of their branching
ratios are much smaller than the experimental upper limits in all of
the parameter space of the $LHT$ model.

\vspace{0.4cm}

\noindent{\bf \large 4. Conclusions and Discussions}

The experimental upper limits of the $LFV$ decay processes
$\tau^-\to \mu^- P_1 P_2$ with $P_1P_2=\pi^+\pi^-$, $K^+K^- $ and
$K^0\bar {K^0}$ have been improved to $\ord(10^{-7})$ at $90\% ~\rm
C.L.$ \cite{14-belle,155}.  Whether these $LFV$ decay processes
exist or not is very important to the neutrino mass problem in the
$SM$. It is well known that the $SM$ does not allow the $LFV$
processes at tree level, while many popular $NP$ models can induce
the $LFV$ processes at tree level or loop level which might make
their branching ratios significantly larger than those predicted by
the $SM$. So the $LFV$ decay processes $\tau^-\to \mu^- P_1 P_2$ are
very suitable for the determination of the free parameters of the
$NP$ models. Studying of these decay processes are very interesting
and needed.

The $TC2$ model and the $LHT$ model are two kinds of the popular
$NP$ models at present. In this paper, we have calculated their
contributions to the branching ratios of the $LFV$ decay processes
$\tau^-\to \mu^- P_1 P_2$. We find that the new particles predicted
by these two $NP$  models can indeed produce significant
contributions to these $LFV$ decay processes. Taking into account
the limits of the relevant experimental data on the free parameters,
we calculate the $BR(\tau^- \to \mu^-P_1 P_2)$, and have the
following conclusions.

i) The $TC2$ model can induce the $LFV$ decay processes $\tau^-\to
\mu^- P_1 P_2$ both at tree level and one loop level, while the
$LHT$ model can only give contributions to these processes at one
loop level. Furthermore, in the case of that the T-odd leptons are
degenerate, the $LHT$ model has no contributions to these $LFV$
decay processes.

ii) For these two $NP$ models, the branching ratios satisfy the
following hierarchy: $Br(\tau^- \to \mu^- \pi^+ \pi^-)$ $>$
$BR(\tau^- \to \mu^- K^+ K^-)$ $\gtrsim$ $BR(\tau^- \to \mu^- K^0
\bar{K}^0)$.

iii) The contributions of the nonuniversal gauge boson $Z'$ at tree
level to the $LFV$ decay processes $\tau^-\to \mu^- \pi^+\pi^-$,
$\tau^-\to \mu^- K^+K^-$ and $\tau^-\to \mu^- K^0\bar {K^0}$ are
larger than those at one loop level by one order of magnitude in
most of the parameter space. However, these values are still not
large enough to be detected by present high energy experiments,
which still need the future experimental verification.

iv) The branching ratios of the $LFV$ decay processes $\tau^-\to
\mu^- P_1 P_2$ generated by the $LHT$ model are much larger than
those generated by the $TC2$ model. For $M_{l^{\tau}_H}=1500~\rm GeV$ and
$f=500~\rm GeV$, the value of the branching ratio $Br(\tau^-\to
\mu^- \pi^+\pi^-)$ can reach $3.14\times 10^{-8}$, which might
approach the upper limit given in Eqs.~(\ref{1}). However, the
values of the branching ratios of the $LFV$ decay processes
$\tau^-\to \mu^- K^+K^-$ and $\tau^-\to \mu^- K^0\bar {K^0}$ are
smaller than $1\times10^{-9}$ in most of parameter space of the
$LHT$ model.


Our calculation can be extended to the $LFV$ decay process
$\tau^-\to e^- P_1 P_2$ by replacing the mass parameter $m_{\mu}$ to
$m_e$. Since the nonuniversal gauge boson $Z'$ treats the first
generation fermions same as those in the second generation, the
coefficient of the coupling $Z'\tau\mu$ approximately equals to that
of the coupling $Z'\tau e$. This feature leads to the fact that the
contribution of the $TC2$ model to the decay $\tau^-\to e^- P_1 P_2$
is nearly the same as that of the decay $\tau^-\to \mu^- P_1 P_2$
channel. For the $LHT$ model, its contributions to the $LFV$ decay
processes $\tau^-\to e^-( \mu^-) P_1 P_2$ can only exist at one
loop. The relevant flavour mixing matrix elements and the masses of
new particles for the $LFV$ decay process $\tau^-\to e^- P_1 P_2$
are different from those of the $LFV$ decay process $\tau^-\to \mu^-
P_1 P_2$, which can make the branching ratios of these two decay
processes different from each other. However, if we neglect these
differences, the value of the branching ratio for the decay process
$\tau^-\to \mu^- P_1 P_2$ should approximately equals to that of the
decay process $\tau^-\to e^- P_1 P_2$.

\vspace{0.3cm}

\noindent{\bf \large Acknowledgments}

\vspace{0.2cm} This work was supported in part by the National
Natural Science Foundation of China under Grants No.10675057 and
10975067, Specialized Research Fund for the Doctoral Program of
Higher Education(SRFDP) (No.200801650002), the Natural Science
Foundation of the Liaoning Scientific Committee(No.20082148), and
Foundation of Liaoning Educational Committee(No.2007T086).

\vspace{0.5cm}

\noindent{\bf \large Appendix}

\vspace{0.3cm}

\noindent{ \bf A. The relevant functions of the hadronic form
factors}

In this appendix we list the hadronic form factors that are related
to the $LFV$ $\tau$ decays $\tau^-\to \mu^- P_1 P_2$. Their explicit
expressions have been given in Ref. \cite{3}, we just put the
related functions as follows:
\begin{eqnarray}
F^{\pi^+ \pi^-}(s) & = & F(s) \, \, \exp \left[ 2 \, Re
\left(\tilde{H}_{\pi \pi}(s) \right) \, + \, Re \left(\tilde{H}_{KK}
(s)\right) \right] ,\\
F^{K^+K^-}(s) & = &F_{\rho}(s)+F_{\omega}(s)+F_{\phi}(s),\\
F^{K^0\bar{K^0}}(s) & = &-F_{\rho}(s)+F_{\omega}(s)+F_{\phi}(s),
\end{eqnarray}
with
\begin{eqnarray}
F(s) & =  & \frac{M_{\rho}^2}{M_{\rho}^2-s-i M_{\rho}
\Gamma_{\rho}(s)} \left[ 1 + \left( \delta \,
\frac{M_{\omega}^2}{M_{\rho}^2} \, - \, \gamma \,
\frac{s}{M_{\rho}^2}
\right) \, \frac{s}{M_{\omega}^2-s- i M_{\omega} \Gamma_{\omega}} \right] \nonumber \\
& &  - \frac{\gamma \, s}{M_{\rho'}^2-s-i M_{\rho'}
\Gamma_{\rho'}(s)} \, \, , \nonumber \\
F_{\rho}(s)&=&\frac{1}{2} \,\frac{M_{\rho}^2}{M_{\rho}^2-s-i
M_{\rho} \Gamma_{\rho}(s)} \, \exp \left[ 2 \, Re
\left(\tilde{H}_{\pi \pi}(s) \right) \, + \, Re
\left(\tilde{H}_{KK} (s)\right) \right]  \, ,\nonumber \\
F_{\omega}(s)& =&    \, \frac{1}{2} \, \left[ \sin^2 \theta_V \,
\frac{M_{\omega}^2}{M_{\omega}^2-s-i M_{\omega} \Gamma_{\omega}} \,
\right] \,   \exp \left[ 3 \,  Re \left(\tilde{H}_{KK} (s)
\,\right)\right]  \, ,\nonumber \\
F_{\phi}(s) &= &    \, \frac{1}{2} \, \left[  \,\cos^2 \theta_V \,
\frac{M_{\phi}^2}{M_{\phi}^2-s-i M_{\phi} \Gamma_{\phi}} \right]
\exp \left[ 3 \,  Re \left(\tilde{H}_{KK} (s) \,\right)
\right]  ,\nonumber\\
\Gamma_{\rho}(s)  &= & \frac{M_{\rho} s}{96 \pi F^2} \left[
\sigma_{\pi}^3(s) \, \theta(\, s \, - \, 4 m_{\pi}^2) \, + \,
\frac{1}{2} \, \sigma_K^3(s) \, \theta( \, s \, - \, 4 m_K^2)
\right] \, ,\nonumber\\
\Gamma_{\rho'}(s) & =&  \Gamma_{\rho'}(M_{\rho'}^2) \,
\frac{s}{M_{\rho'}^2} \, \left( \frac{\sigma_{\pi}^3(s) \, \, + \,
\frac{1}{2} \, \sigma_K^3(s) \, \theta( \, s \, - \, 4
m_K^2)}{\sigma_{\pi}^3(M_{\rho'}^2) \,  + \,  \frac{1}{2} \,
\sigma_K^3(M_{\rho'}^2) \, \theta( \, s \, - \, 4 m_K^2)} \right)
\theta(\, s \, - \, 4 m_{\pi}^2)  \, .
\end{eqnarray}
where $\sigma_P(s)  =  \sqrt{1-4 \frac{m_P^2}{s}}$, and the other
definitions are:
\begin{eqnarray} \label{eq:functions}
\beta & = & \frac{\Theta_{\rho \omega}}{3 M_{\rho}^2} \; ,\nonumber \\
\gamma & = & \frac{F_V G_V}{F^2} \left( 1+ \beta \right) - 1 \; , \nonumber \\
\delta & = & \frac{F_V G_V}{F^2} - 1 \; , \nonumber \\
\tilde{H}_{PP}(s) & = & \frac{s}{F^2}  M_{P}(s)  \; , \nonumber \\
M_P(s) & = & \frac{1}{12} \left( 1 - 4 \frac{m_P^2}{s} \right) \,
J_P(s) \,
- \, \frac{k_P(M_{\rho})}{6} \, + \, \frac{1}{288 \pi^2} \; , \nonumber \\
J_P(s) & = & \frac{1}{16 \pi^2} \left[ \sigma_P(s) \, \ln
\frac{\sigma_P(s) -1}{\sigma_P(s) + 1}
+ 2 \right] \; , \nonumber \\
k_P(\mu) & = & \frac{1}{32 \pi^2} \left( \ln \frac{m_P^2}{\mu^2}+1
\right) \; .
\end{eqnarray}
The contribution of the isospin breaking $\rho - \omega$ mixing
$\Theta_{\rho \omega} = -3.3 \times 10^{-3} \, \mbox{GeV}^2$, and
the asymptotic constraint on the $N_C \rightarrow \infty$ vector
form factor indicates $F_V G_V \simeq F^2=F_{\pi}^2$. The mixing
between the octet and singlet vector components employed in the
construction of the $I=0$ component of the kaon vector form factors
is defined by~:
\begin{equation}
\left(  \begin{array}{c}
\phi \\
\omega
\end{array} \right)\, = \,  \left(
\begin{array}{cc}
\cos \theta_V & - \sin \theta_V \\
\sin \theta_V &  \cos \theta_V
\end{array}  \right) \;  \left(
\begin{array}{c}
v_8 \\
v_0
\end{array} \right)\, ,
\end{equation}
and the ideal mixing $\theta_V = 35^{\circ}$ was used.

\vspace{0.3cm}

\noindent{ \bf B. The relevant functions in the $TC2$ model}

In the framework of $TC2$ model, the Inami-Lim functions that are
used in our calculation are given as following.
\begin{eqnarray}
F_1(x)&=&\frac{1}{8}\left[\frac{x^2lnx}{(x-1)^2}-\frac{2xlnx}{(x-1)^2}+\frac{x}{x-1}\right];\\
F_2(x)&=&-\frac{1}{4}\left[\frac{x}{x-1}-\frac{x ln
x}{(x-1)^2}\right];\\
F_3(x)&=&\frac{1}{32}\left[\frac{x^2lnx}{(x-1)^2}-\frac{x}{x-1}
-\frac{x\gamma_E}{2}-x ln 4\pi-\frac{3x^2}{8}\right.\nonumber\\
&&\left.+\frac{x^4lnx}{4(x-1)^2}-\frac{x^2}{4(x-1)}\right];\\
F_4(x)&=&-\frac{x}{16}\left[\frac{-1}{4(x-1)}+\frac{3}{4(x-1)^2}
+\frac{3}{2(x-1)^3}-\frac{3xlnx}{(x-1)^4}\right].
\end{eqnarray}

\vspace{0.3cm}

\noindent{ \bf C. The relevant functions in the $LHT$ model}

In this appendix we enumerate the functions related our calculation
of the $LFV$ $\tau$ decays $\tau^-\to \mu^- P_1 P_2$ in the $LHT$
model, which have been discussed in Ref. \cite{201}.
\begin{eqnarray}
S_{odd}(x)&=&\frac{x^2-2x+4}{(1-x)^2}lnx+\frac{7-x}{2(1-x)},\\
F^{u \bar{u}}\left(y_{i}, z; W_{H}\right) &=& \frac{3}{2} y_{i}
- F_{5}\left(y_{i}, z\right) - 7 F_{6}\left(y_{i}, z\right) - 9 U\left(y_{i}, z\right)\,, \\
F^{d\bar d}\left(y_{i}, z; W_{H}\right) &=& \frac{3}{2} y_{i} -
F_{5}\left(y_{i}, z\right)- 7 F_{6}\left(y_{i}, z\right) + 3
U\left(y_{i}, z\right)\,,\\
F_{5}\left(y_{i}, z\right) &=& \frac{y_{i}^{3} \log y_{i}}{\left(1-y_{i}\right) \left(z-y_{i}\right)}
+ \frac{z^{3} \log z}{\left(1-z\right) \left(y_{i}-z\right)}\,, \\
F_{6}\left(y_{i}, z\right) &=& -\left[\frac{y_{i}^{2} \log y_{i}}{\left(1-y_{i}\right) \left(z
-y_{i}\right)} + \frac{z^{2} \log z}{\left(1-z\right) \left(y_{i}-z\right)}\right]\,, \\
U\left(y_{i}, z\right) &=& \frac{y_{i}^{2} \log
y_{i}}{\left(y_{i}-z\right) \left(1-y_{i}\right)^{2}} + \frac{z^{2}
\log z}{\left(z-y_{i}\right) \left(1-z\right)^{2}} +
\frac{1}{\left(1-y_{i}\right) \left(1-z\right)}\,\\
G\left(y_{i}, z; Z_{H}\right) &=&  -\frac{3}{4} U\left(y_i, z\right) \,,\\
G_{1}\left(y_{i}^{\prime},z^{\prime}; A_{H}\right) &=&
\frac{1}{25 a} G\left(y_{i}^{\prime}, z^{\prime}; Z_{H}\right) \,,\\
G_{2}\left(y_{i}, z; \eta\right) &=& -\frac{3}{10 a}
\left[\frac{y_{i}^{2} \log y_{i}}{\left(1-y_{i}\right)
\left(\eta-y_{i}\right) \left(y_{i}-z\right)} \right.\nonumber \\
&& +  \left. \frac{z^{2} \log z}{\left(1-z\right)
\left(\eta-z\right) \left(z-y_{i}\right)} + \frac{\eta^{2} \log
\eta}{\left(1-\eta\right) \left(y_{i}-\eta\right)
\left(\eta-z\right)}\right]\,,\\
D_0'(x)&=&-\frac{3x^3-2x^2}{2(x-1)^4}lnx+\frac{8x^3+5x^2-7x}{12(x-1)^3},\\
E_0'(x)&=&\frac{3x^2}{2(x-1)^4}lnx+\frac{x^3-5x^2-2x}{4(x-1)^3}.
\end{eqnarray}

\null
\end{document}